\documentclass[12pt]{spieman}  
\usepackage{amsmath,amsfonts,amssymb}
\usepackage{graphicx}
\usepackage{setspace}
\usepackage{tocloft}
\usepackage{epsfig}

\title{Memristive model of hysteretic field emission from carbon nanotube arrays}

\author[a]{D. V. Gorodetskiy}
\author[a]{A. V. Gusel'nikov}
\author[b,c]{S. N. Shevchenko}
\author[a,d]{M. A. Kanygin}
\author[a,d]{A.\thinspace V.\thinspace Okotrub}
\author[a,e,*]{Y. V. Pershin}
\affil[a]{Nikolaev Institute of Inorganic Chemistry SB RAS, 630090 Novosibirsk, Russia}
\affil[b]{B. Verkin Institute for Low Temperature Physics and Engineering, Kharkov, Ukraine}
\affil[c]{V. Karazin Kharkov National University, Kharkov, Ukraine}
\affil[d]{Novosibirsk State University, Novosibirsk 630090, Russia}
\affil[e]{Department of Physics and Astronomy, University of South Carolina, Columbia, SC 29208 USA}

\cftpagenumbersoff{figure}
\cftpagenumbersoff{table}
\begin{document}
\maketitle

\begin{abstract}
Some instances of electron field emitters are characterized by frequency-dependent hysteresis in their current-voltage characteristics.
We argue that such emitters can be classified as memristive systems and introduce a general framework to describe their response. As a specific example
of our approach, we consider field emission from a carbon nanotube array. Our experimental results demonstrate a low-field hysteresis, which is likely caused by an electrostatic alignment of some of the nanotubes in the applied field. We formulate a memristive model of such phenomenon whose results are in agreement with the experimental results.
\end{abstract}

\keywords{memristor; field emission; carbon nanotubes; hysteresis; Fowler-Nordheim equation}

{\noindent \footnotesize\textbf{*}Y. V. Pershin,  \linkable{pershin@physics.sc.edu} }

\begin{spacing}{2}   

\section{Introduction}

In the past two decades, the study of field emission from carbon nanomaterials has attracted significant attention \cite{saito2010carbon,Eletskii10a,Li15a}. Potential applications for field-emission devices include  X-ray tubes \cite{Hwan07a}, high-luminance light sources \cite{Saito200230} and field-emission displays \cite{Choi99a}. While most of the attention has been focused on field emission from carbon nanotube (CNT) arrays, the field emission from carbon yarns, fibers and graphene flakes has also been reported.

The field emission properties of carbon materials depend on many experimental parameters. While, ideally, it is often expected that the current-voltage curves follow the Fowler-Nordheim theory \cite{FN28a}, some publications report significant deviations from such an ideal response including a hysteresis in current-voltage characteristics \cite{Lim01a,Park20052094,Sveningsson05a,Li08c,Okotrub09a,Chen11a,Cahay14a,Kleshch15a,popov2015comparison} and self-oscillations \cite{Barois13a}. The origin of the hysteresis is typically associated with either adsorption/desorption processes \cite{Lim01a,Park20052094,Li08c,Okotrub09a,Chen11a}, electrostatic alignment of some constituents of the system \cite{Cahay14a,Kleshch15a}, or temperature effects \cite{Sveningsson05a}. As of now, there is no common theory that would account for such dynamic processes in the field emission.

\begin{figure*}[t]
\centering
\includegraphics[width=0.85\textwidth]{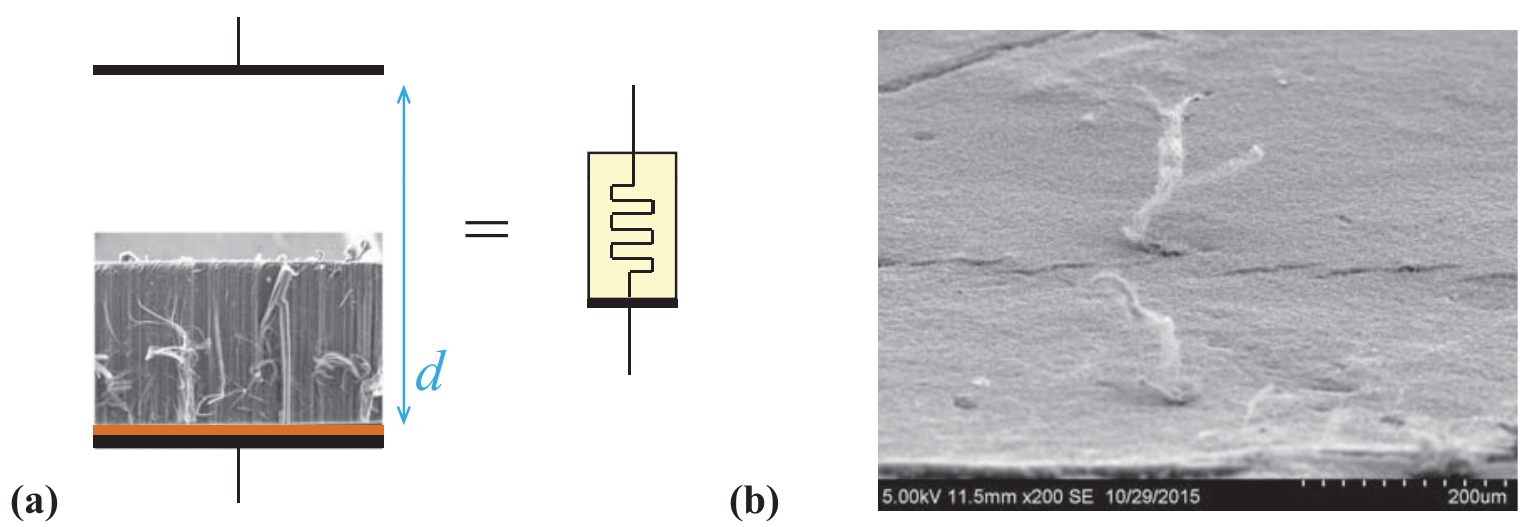}
\caption{ \textbf{(a)} CNT array field emission device comprises a memristive system. The CNT array is attached to the cathode (at the bottom) separated by a distance $d$ from the anode (at the top). The circuit symbol of memristor is shown to the right. As nonlinear resistors with memory (see Eq. (\ref{eq1a}) definition), memristors are capable of describing strongly asymmetric (diode-like) response with memory of our  field-emission devices. \textbf{(b)} SEM image showing the top surface of the CNT array field emission device.}
\label{fig:fig1}
\end{figure*}

The purpose of the present article is to develop a link between hysteretic field-emission devices and a general class of memristive devices and systems (resistors with memory) \cite{chua76a} (see Fig. \ref{fig:fig1}(a)). This goal is approached both experimentally and theoretically. Experimentally, it is shown that the field emission from carbon nanotube arrays exhibits a frequency-dependent hysteresis in the form of pinched hysteresis loops, which are the most pronounced feature of memristive devices and systems \cite{chua76a}. Theoretically, we introduce a general formalism of memristive systems and develop a specific theoretical model accounting for our experimental observations.

We note that memristive systems~\cite{chua76a} are
particular types of memory circuit elements \cite{diventra09a,diventra09b}.
There are two kinds of memristive systems: voltage-controlled and
current-controlled ones~\cite{chua76a}. The voltage-controlled memristive
systems are defined by the equations
\begin{eqnarray}
I(t) &=&R^{-1}\left( x,U,t\right) U(t),  \label{eq1a} \\
\dot{x} &=&f\left( x,U,t\right) ,  \label{eq2a}
\end{eqnarray}%
where $U(t)$ and $I(t)$ denote the voltage and current across the device, $R$
is the memristance (memory resistance), $x=\{x_{i}\}$ is a set of $n$ state
variables describing the internal state of the system, and $f$ is an $n$%
-dimensional vector function. Current-controlled memristive systems are
defined in a similar way \cite{chua76a}.

\begin{figure*}[tbp]
\centering
\includegraphics[width=0.85\textwidth]{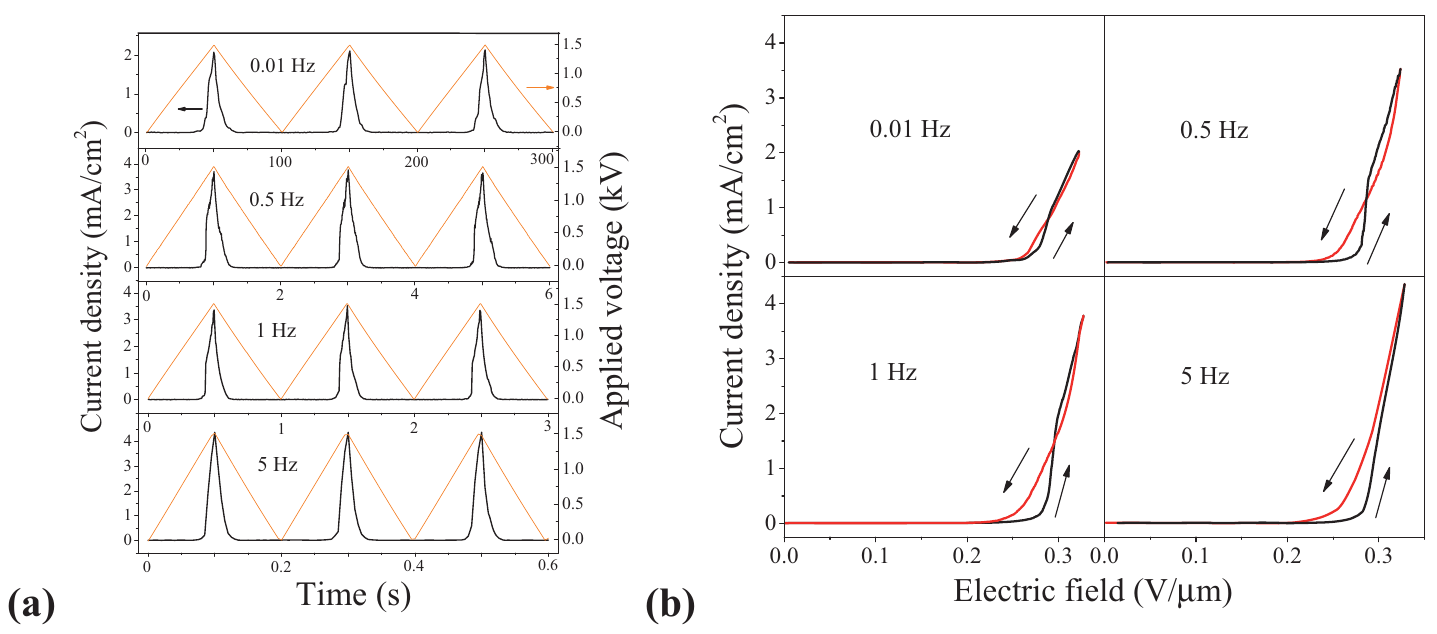}
\caption{\textbf{(a)} Applied voltage and field emission current density as functions of time.
\textbf{(b)} Current-voltage characteristics of field emission at different frequencies of the input signal.}
\label{fig:fig2}
\end{figure*}

The importance of Eqs. (\ref{eq1a}) and (\ref{eq2a}) lies with the fact that
they represent a unified framework to describe various memristive
devices \cite{pershin11a}. Using these equations, one can abstract from
specific details of device realization, thus describing devices based on
different physical principles \cite{pershin11a} on the same foot \cite%
{diventra11a}. Therefore, it is not surprising that many unlike memristive
devices share several features \cite{chua76a}. The best known is
the hysteresis, which is most pronounced at frequencies of the applied
voltage that are comparable to those of internal processes leading to
memory. At much higher frequencies and sometimes at much lower frequencies,
the hysteresis is significantly reduced.

In the following part of this article, the hysteretic behavior of field
emission from CNT arrays is explored experimentally, focusing precisely on
how the frequency of the applied input affects the hysteresis in
current-voltage response.

\section{Experimental details and results}

The method of fabrication of CNT
arrays is similar to the one in Ref.~[\citeonline{fedoseeva2015field}]. Our CVD
reactor consists of a quartz tube (5 cm in diameter) placed inside a
high-temperature furnace. The precision of regulation for the furnace temperature
is about 0.1$^{\circ }$C. Vertically aligned CNT arrays were grown by an
aerosol-assisted catalytic chemical vapor deposition on Si (100) substrates
with lateral dimensions of $10\times 10$ mm$^{2}$. A reaction mixture (2\%
of ferrocene (Fe(C$_{5}$H$_{5}$)$_{2}$) in toluene (C$_{6}$H$_{5}$CH$_{3}$))
was dispersed via an injector located in $\sim 200^{\circ }$C zone at a rate
of 15 ml/hour. High purity Ar at atmospheric pressure was used to transport the
vaporized reaction mixture to the synthesis zone (the central part of the
quartz tube). The pyrolysis was performed at 800$^{\circ }$C. The reaction
mixture was supplied for 40 minutes. The argon
gas flow rate of 600 sccm was kept constant for an additional 15 minutes. After
that, the reactor was cooled to room temperature.

Fig. \ref{fig:fig1}(b) shows an SEM image of a selected sample, which demonstrates a
stable and well-defined hysteresis. We note that
detailed examination of CNT array surface (Fig. \ref{fig:fig1}b) reveals
bulging bundles of CNTs with a height of about 150 $\mu$m. Inside
the bundles, CNTs are parallel relative to each other. These bundles are
inclined relative to the normal.


The field emission from CNT arrays was studied in a diode regime in a vacuum
chamber using a home-made set-up. In the experiment, a Si substrate with
CNTs on top was attached from the bottom to a negative electrode (located on
a sample manipulator) by carbon conductive tape. The sample manipulator was
inserted into the vacuum chamber, whose pressure was set to about $5\cdot
10^{-2}$ Pa. A triangle voltage waveform (from 0 V to 1.5 kV) of several
frequencies in the range from 0.01 Hz to 5 Hz was applied across  an anode
and cathode. The distance $d$ between the flat anode and top surface of CNT
array was set to approximately $5$ mm (see Fig. \ref{fig:fig1}(a)).

Fig. \ref{fig:fig2} presents some typical results obtained from measurements. According to Fig. \ref{fig:fig2}(a), the system response is stable in different cycles. Importantly, the current density peaks are asymmetric, which indicates some changes in the system. The current-voltage curves shown in Fig. \ref{fig:fig2}(b) are hysteretic in nature. We would like to point out that both the hysteresis size and shape depend on the voltage frequency applied. This is a common feature of memristive systems (see review paper [\citeonline{pershin11a}] for examples). Moreover, according to Fig. \ref{fig:fig2}, the field emission is observed at a quite low onset field of about 0.25 V/$\mu $m.


\begin{figure}[t]
\centering
\includegraphics[width=9cm]{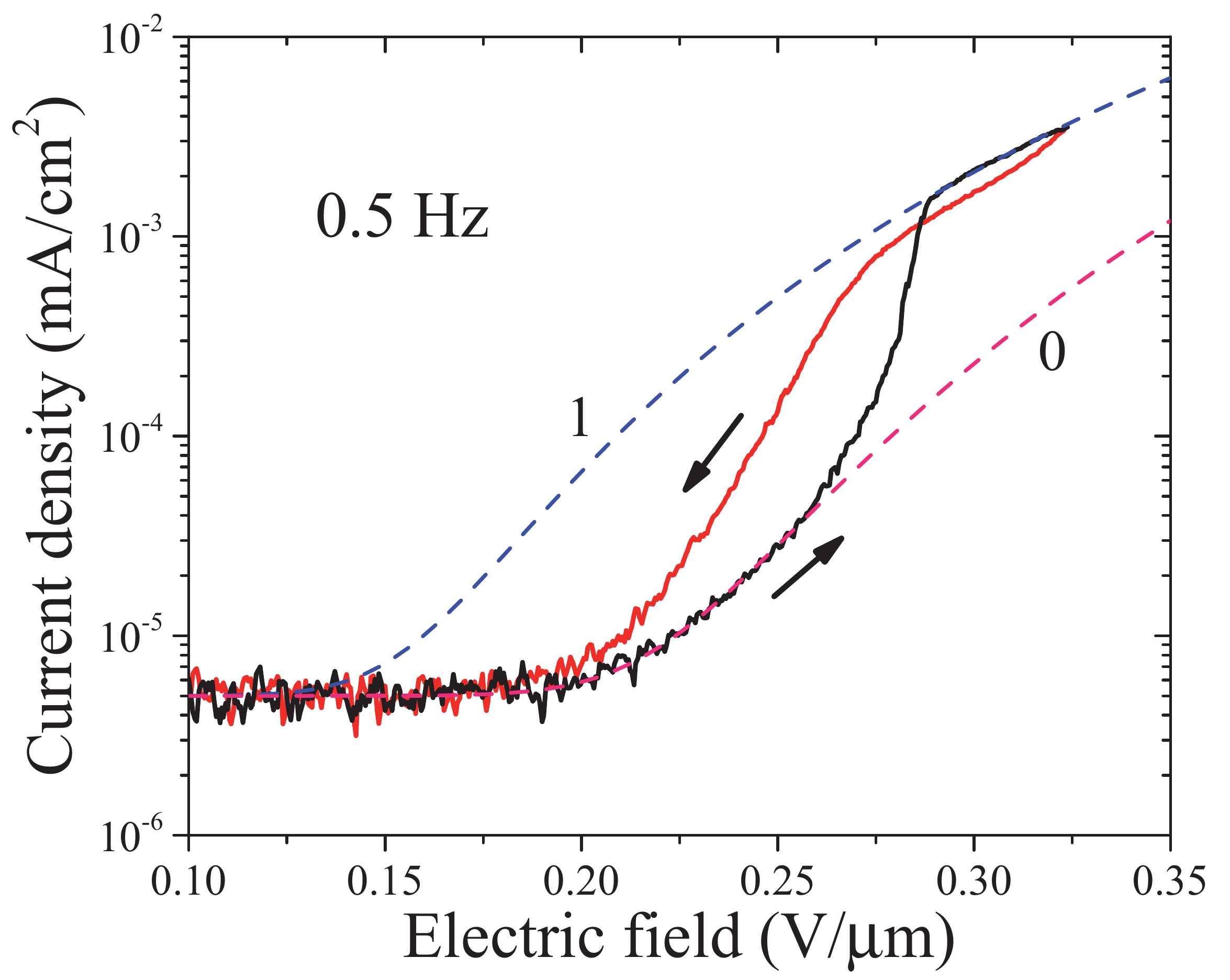}
\caption{ Fitting 0.5 Hz loop in Fig. \protect\ref{fig:fig2}%
(b) with Fowler-Nordheim equation. The direction along the loop is indicated
by arrows. The numbers 0 and 1 on the plot denote the fitting curves $J_{0(1)}$.}
\label{fig:fig3}
\end{figure}

\section{Model}

In order to better understand the system response, we fit
some sections of 0.5 Hz loop (top right plot in Fig. \ref{fig:fig2}(b)) with
the Fowler-Nordheim equation \cite{FN28a}, also taking into account the offset of the amplifier (in-house built)\,$J_{\mathrm{off}}$
\begin{equation}
J_{i}=A_{i}E_0^{2}\exp \left( -\frac{B_{i}}{E_0}\right) +J_{_{\mathrm{off}}}.
\label{eq:FN}
\end{equation}%
Here, $J_{i}$ is the current density, $E_0=U/d$ is the average electric field, and the index $i=0,1$ defines different
fitting curves related to the two
sections of the loop (see Fig. \ref{fig:fig3}). The fitting results shown in Fig. \ref{fig:fig3}
demonstrate that both low- and high-field regions of up- (and down-)sweeps
follow the Fowler-Nordheim formula very closely. There are also transition
regions (from 0.26 V/$\mu $m to 0.29 V/$\mu $m in the up-sweep and from 0.27
V/$\mu $m to 0.24 V/$\mu $m in the down-sweep) deviating from Eq. (\ref%
{eq:FN}) behavior. The same widths of transition regions indicate their
common origin.

We attribute the above-mentioned transition regions to morphological changes in
the system \cite{Cahay14a,Kleshch15a}. Indeed, surfaces of our CNT arrays are
not perfect (see Fig. \ref{fig:fig1}(b)): there are some CNTs (and their bundles)
rising above the surface. These CNTs/bundles form effective springs stretchable by
the applied field. When these springs are stretched, the emission current increases if
the new configuration offers better conditions for the field emission (e.g. stronger
local electric field due to closer distance to the anode).
It seems likely that these longer CNTs/bundles were nucleated at the initial
stage of CNT growth due to, for instance, local surface contamination.


\begin{figure}[t]
\centering
\includegraphics[width=9cm]{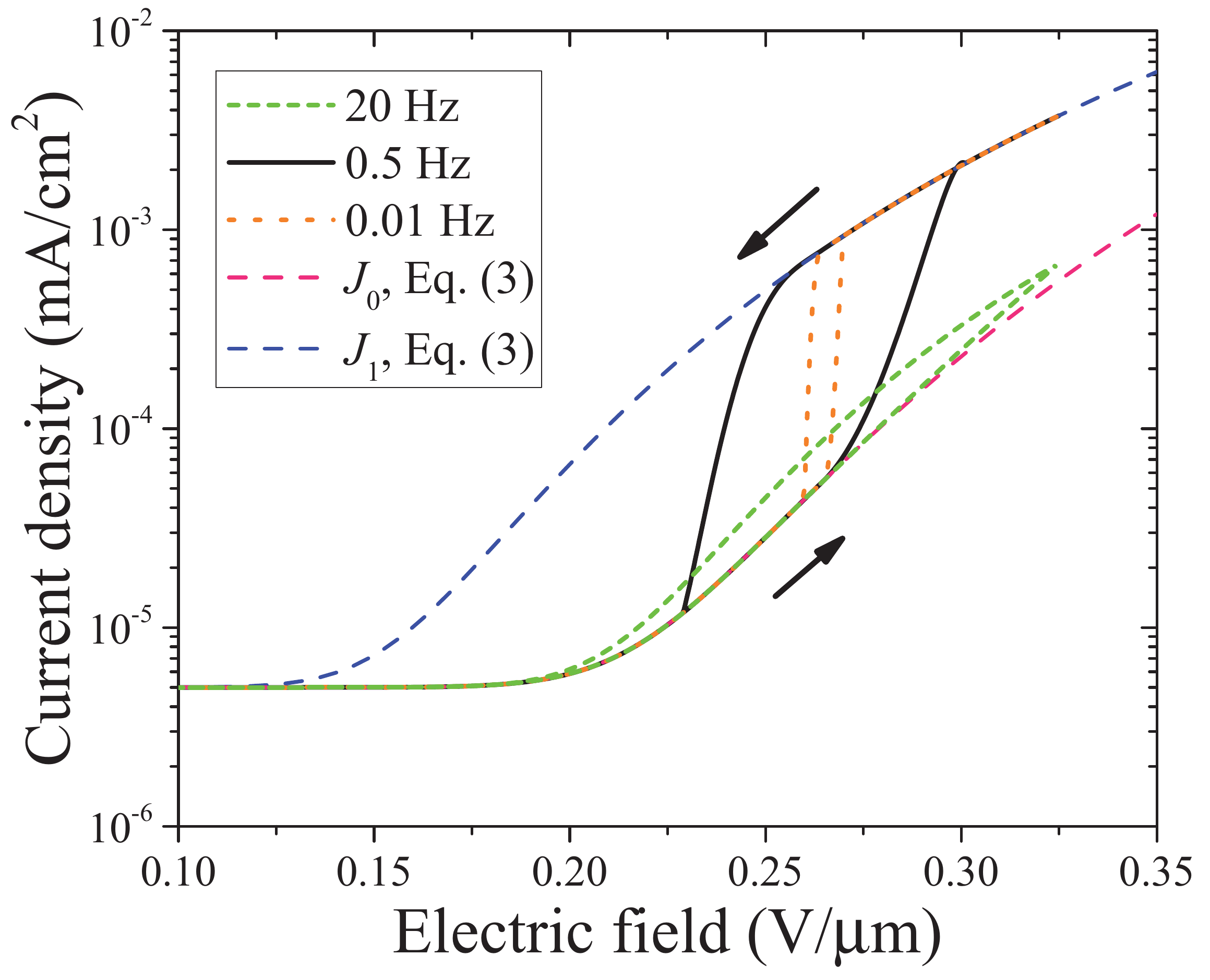}
\caption{ Hysteresis curves found with Eqs. (\protect\ref%
{eq:FN1}) and (\protect\ref{eq:z}). This plot was obtained using the
following parameter values: $E_{\mathrm{c}}=0.265$ V/$\protect\mu $m, $%
\protect\alpha =500$ $\protect\mu $m/(V$\cdot $s), $A_{0}=33.8$~(mA/cm$^2$)/(V/$\mu$m)$^{2}$,
$A_{1}=5.40$~(mA/cm$^2$)/(V/$\mu$m)$^{2}$, $B_{0}=2.85$ V/$\mu$m, $B_{1}=1.63$ V/$\mu$m,
and $J_{\textnormal{off}}=5\cdot 10^{-6}$ mA/cm$^2$.}
\label{fig:fig4}
\end{figure}

Additionally, taking a closer look at Fig. %
\ref{fig:fig2}(b) one can notice that on average the current density
decreases with frequency decrease. This reduction might be related to
adsorption/desorption processes on CNTs \cite{Lim01a,Park20052094,Li08c,Okotrub09a,Chen11a}.
Moreover, CNT heating may also play a certain role in hysteretic response. However,
the increase of current density with temperature is expected.
In what follows we focus mainly on the low field response, considering only
physically reasonable morphological mechanism explaining sharp steps (transition regions) in hysteresis.



In order to formulate its
mathematical model in the memristive form (Eqs. (\ref{eq1a}), (\ref{eq2a})),
it is first necessary to identify the relevant internal state variables. For
the sake of simplicity, we use a single dimensionless variable $z$
characterizing the degree of alignment of effective emitters. In
particular, it is assumed that $z=0$ corresponds to the completely unaligned
state (the equilibrium configuration at $U=0$), and $z=1$ corresponds to the
aligned one (the equilibrium configuration at very large fields).

The Fowler-Nordheim equation with $z$-dependent coefficients is employed as
the first equation defining the memristive system (Eq. (\ref{eq1a})):
\begin{equation}
J(t)=A(z)E_0(t)^{2}\exp \left( -\frac{B(z)}{E_0(t)}\right) +J_{_{\mathrm{off}}}.
\label{eq:FN1}
\end{equation}%
In the simplest linear approximation one may take $A(z)=A_{0}(1-z)+A_{1}z$
and $B(z)=B_{0}(1-z)+B_{1}z$, where $A_{i}$ and $B_{i}$ are the fitting
constants introduced in Eq. (\ref{eq:FN}). In this way, the model of
current-voltage curve is based on the first fitting function at small
voltages and the second one at large voltages.

The current-voltage curves presented in Fig. \ref{fig:fig2}(b) suggest that
there is a critical field $E_{\mathrm{c}}$ for CNT alignment. Using a linear
approximation for $f(U)$, we write Eq. (\ref{eq2a}) as
\begin{equation}
\frac{\mathnormal{d}z}{\mathnormal{d}t}=\alpha (E_{0}-E_{\mathrm{c}}),
\label{eq:z}
\end{equation}%
where $\alpha $ is a constant, and $z$ is the variable
constrained to the interval between 0 and 1. According to the above
equation, $z$ increases (decreases) when the electric field is above (below)
$E_{\mathrm{c}}$.

Fig. \ref{fig:fig4} presents results of our simulations based on Eqs. (\ref%
{eq:FN1}) and (\ref{eq:z}). This plot demonstrates a larger hysteresis at $%
\nu=0.5$ Hz, and smaller ones at lower and higher frequencies. We emphasise
that even with a very simple form of the right-hand side of Eq. (\ref{eq:z}%
), our theoretical curves are in qualitative agreement with the
experiment. This indicates suitability of the selected modeling approach.

Finally, we
would like to note that temperature effects can be taken into
account by replacing $A_i$ in Eq.~(\ref{eq:FN}) by a temperature-dependent factor
defined by the Murphy-Good theory \cite{Murphy56a,Sheshin01}. The effective CNT temperature
can be considered as another internal state variable described by the heat-balance equation.
Such calculations for high temperatures
also result in the hysteretic current-voltage characteristics.
These calculations are related rather to the strong field limit
and will be reported elsewhere.

\emph{Conclusion.}-- We have demonstrated that current-voltage characteristics
of field emission from CNT arrays can be hysteretic. We argue that the leading mechanism for the
low-field hysteresis is the morphological one, which is associated with aligning CNTs in the
electric field. A memristive model of hysteresis has been developed. It is anticipated that the
memristive properties of such hysteresis could result in unexpected applications.


\acknowledgments
The work was supported by Russian Scientific Foundation grant 15-13-20021.

\bibliography{memcapacitor}   

\begin{thebibliography}{10}

\bibitem{saito2010carbon}
Y.~Saito, Ed., {\em Carbon Nanotube and Related Field Emitters: Fundamentals
  and Applications}, Wiley-VCH  (2010).

\bibitem{Eletskii10a}
A.~V. Eletskii, ``Carbon nanotube-based electron field emitters,'' {\em
  Physics-Uspekhi} {\bf 53}, 863  (2010).

\bibitem{Li15a}
Y.~Li, Y.~Sun, and J.~T.~W. Yeow, ``Nanotube field electron emission:
  principles, development, and applications,'' {\em Nanotechnology} {\bf 26},
  242001  (2015).

\bibitem{Hwan07a}
S.~H. Heo, A.~Ihsan, and S.~O. Cho, ``Transmission-type microfocus {X}-ray tube
  using carbon nanotube field emitters,'' {\em Appl. Phys. Lett.} {\bf 90},
  183109  (2007).

\bibitem{Saito200230}
Y.~Saito, K.~Hata, A.~Takakura, J.~Yotani, and S.~Uemura, ``Field emission of
  carbon nanotubes and its application as electron sources of ultra-high
  luminance light-source devices,'' {\em Physica B: Condensed Matter} {\bf
  323}, 30--37  (2002).

\bibitem{Choi99a}
W.~B. Choi, D.~S. Chung, J.~H. Kang, H.~Y. Kim, Y.~W. Jin, I.~T. Han, Y.~H.
  Lee, J.~E. Jung, N.~S. Lee, G.~S. Park, and J.~M. Kim, ``Fully sealed,
  high-brightness carbon-nanotube field-emission display,'' {\em Appl. Phys.
  Lett.} {\bf 75}, 3129--3131  (1999).

\bibitem{FN28a}
R.~H. {Fowler} and L.~{Nordheim}, ``{Electron Emission in Intense Electric
  Fields},'' {\em Proc. R. Soc. Lond. A} {\bf 119}, 173--181  (1928).

\bibitem{Lim01a}
S.~C. Lim, H.~J. Jeong, Y.~S. Park, D.~S. Bae, Y.~C. Choi, Y.~M. Shin, W.~S.
  Kim, K.~H. An, and Y.~H. Lee, ``Field-emission properties of vertically
  aligned carbon-nanotube array dependent on gas exposures and growth
  conditions,'' {\em J. Vac. Sci. Technol. A} {\bf 19}, 1786--1789  (2001).

\bibitem{Park20052094}
K.~H. Park, S.~Lee, and K.~H. Koh, ``Growth and high current field emission of
  carbon nanofiber films with electroplated {Ni} catalyst,'' {\em Diamond and
  Related Materials} {\bf 14}, 2094 -- 2098  (2005).

\bibitem{Sveningsson05a}
M.~Sveningsson, K.~Hansen, K.~Svensson, E.~Olsson, and E.~E.~B. Campbell,
  ``Quantifying temperature-enhanced electron field emission from individual
  carbon nanotubes,'' {\em Phys. Rev. B} {\bf 72}, 085429  (2005).

\bibitem{Li08c}
C.~Li, G.~Fang, X.~Yang, N.~Liu, Y.~Liu, and X.~Zhao, ``Effect of adsorbates on
  field emission from flame-synthesized carbon nanotubes,'' {\em J. Phys. D:
  Appl. Phys.} {\bf 41}, 195401  (2008).

\bibitem{Okotrub09a}
A.~V. Okotrub, A.~G. Kurenya, A.~V. Gusel'nikov, A.~G. Kudashov, L.~G.
  Bulusheva, A.~S. Berdinskii, Y.~A. Ivanova, D.~K. Ivanov, E.~A. Strel'tsov,
  D.~Fink, A.~V. Petrov, and E.~K. Belonogov, ``The field emission properties
  of carbon nanotubes and {SiC} whiskers synthesized over {Ni} particles
  deposited in ion tracks in {SiO2},'' {\em Nanotechnologies in Russia} {\bf
  4}, 627--633  (2009).

\bibitem{Chen11a}
J.~Chen, J.~Li, J.~Yang, X.~Yan, B.-K. Tay, and Q.~Xue, ``The hysteresis
  phenomenon of the field emission from the graphene film,'' {\em Appl. Phys.
  Lett.} {\bf 99}, 173104  (2011).

\bibitem{Cahay14a}
M.~Cahay, P.~T. Murray, T.~C. Back, S.~Fairchild, J.~Boeckl, J.~Bulmer,
  K.~K.~K. Koziol, G.~Gruen, M.~Sparkes, F.~Orozco, and W.~O'Neill,
  ``Hysteresis during field emission from chemical vapor deposition synthesized
  carbon nanotube fibers,'' {\em Appl. Phys. Lett.} {\bf 105}, 173107  (2014).

\bibitem{Kleshch15a}
V.~I. Kleshch, D.~A. Bandurin, A.~S. Orekhov, S.~T. Purcell, and A.~N.
  Obraztsov, ``Edge field emission of large-area single layer graphene,'' {\em
  Applied Surface Science} {\bf 357, Part B}, 1967 -- 1974  (2015).

\bibitem{popov2015comparison}
E.~O. Popov, A.~G. Kolosko, S.~V. Filippov, P.~A. Romanov, and I.~L. Fedichkin,
  ``Comparison of the data about thin {IVC} structure of multi-tip field
  emitters using a high voltage scanning method in different power supply mode
  and the data of the mass spectrometer analysis,'' in {\em Vacuum
  Nanoelectronics Conference {(IVNC)}, 2015 28th International},  18--19, IEEE
  (2015).

\bibitem{Barois13a}
T.~Barois, S.~Perisanu, P.~Vincent, S.~T. Purcell, and A.~Ayari, ``Role of
  fluctuations and nonlinearities on field emission nanomechanical
  self-oscillators,'' {\em Phys. Rev. B} {\bf 88}, 195428  (2013).

\bibitem{chua76a}
L.~O. Chua and S.~M. Kang, ``Memristive devices and systems,'' {\em Proc.
  {IEEE}} {\bf 64}, 209--223  (1976).

\bibitem{diventra09a}
M.~{Di Ventra}, Y.~V. Pershin, and L.~O. Chua, ``Circuit elements with memory:
  Memristors, memcapacitors, and meminductors,'' {\em Proc. {IEEE}} {\bf
  97}(10), 1717--1724  (2009).

\bibitem{diventra09b}
M.~{Di Ventra}, Y.~V. Pershin, and L.~O. Chua, ``Putting memory into circuit
  elements: memristors, memcapacitors and meminductors,'' {\em Proc. {IEEE}}
  {\bf 97}(8), 1371--1372  (2009).

\bibitem{pershin11a}
Y.~V. Pershin and M.~Di~Ventra, ``Memory effects in complex materials and
  nanoscale systems,'' {\em Advances in Physics} {\bf 60}, 145--227  (2011).

\bibitem{diventra11a}
M.~Di~Ventra and Y.~V. Pershin, ``Memory materials: a unifying description,''
  {\em Materials Today} {\bf 14}, 584  (2011).

\bibitem{fedoseeva2015field}
Y.~V. Fedoseeva, L.~Bulusheva, A.~Okotrub, M.~Kanygin, D.~Gorodetskiy,
  I.~Asanov, D.~Vyalikh, A.~Puzyr, and V.~Bondar, ``Field emission luminescence
  of nanodiamonds deposited on the aligned carbon nanotube array,'' {\em
  Scientific Reports} {\bf 5}, 9379  (2015).

\bibitem{Murphy56a}
E.~L. Murphy and R.~H. Good, ``Thermionic emission, field emission, and the
  transition region,'' {\em Phys. Rev.} {\bf 102}, 1464--1473  (1956).

\bibitem{Sheshin01}
E.~P. Sheshin, {\em Surface structure and electron field emission properties of
  carbon materials}, MFTI, Moscow  (2001).

\end{thebibliography}
\bibliographystyle{spiejour}   





\end{spacing}
\end{document}